\documentclass[doublecol]{epl2} 
\usepackage{graphicx}
\usepackage{epsfig}
\usepackage{epstopdf}
\usepackage{dcolumn}
\usepackage{bm}

\title{Bond Orientational Order Parameters in the Crystalline Phases of the Classical Yukawa-Wigner Bilayers.}
\shorttitle{Crystalline Phases of the Classical Yukawa-Wigner Bilayers.} 

\author{Martial Mazars\inst{1}}
\shortauthor{M. Mazars }

\institute{\inst{1} Laboratoire de Physique Th\'eorique (UMR 8627), Universit\'e de Paris Sud XI, B\^atiment 210, 91405 Orsay Cedex, FRANCE \\}

\abstract{We present a study of the structural properties of the crystalline phases for a planar bilayer of particles interacting via repulsive Yukawa potentials in the weak screening region. The study is done with Monte Carlo computations and the long ranged contributions to energy are taken into account with the Ewald method for quasi-two dimensional systems. Two first order phase transitions (fluid-solid and solid-solid) and one second order transition (solid-solid) are found when the surface density is varied at constant temperature. A particular attention is pay to the characteristics of the crystalline phases by the analysis of bond orientational order parameters and center-to-center correlations functions.}
\begin{document}
\maketitle
{\bf Introduction} - Two dimensional crystals of charged particles were first observed on a monolayer of electron adsorbed at the surface of superfluid helium \cite{Grimes:79} ;  multilayers of charged ions in laser-beam-cooled trapped plasma \cite{Mitchell:98}, dusty plasma \cite{Thomas:94,Chu:94,Zuzic:00,Fortov:05,Monarkha:03} and in colloidal suspensions \cite{Pieranski:83,Zahn:99} have been observed as well. In particular, bilayer systems have benefited from a large number of theoretical \cite{Goldoni:96,Valtchinov:97,Messina:03} and numerical \cite{Totsuji:97,Schweigert:99,Weis:01} studies.\\
In colloidal and dusty plasma systems, the interactions between particles are quite well approximated by Yukawa potentials of the form 
\begin{equation}
\label{YukE}
\displaystyle V(r)= \frac{Q^2\exp(-r/\lambda_D)}{r}
\end{equation}
where $Q$ is the charge carried by the particles, $r$ the distance between particles and $\lambda_D$ the Debye screening length. In ref.\cite{Messina:03}, the phase diagram of the Yukawa bilayer at $T=0$ has been computed and six different crystal phases have been identified. Previous numerical simulations on bilayers systems have adressed either the unscreened Coulomb interaction ($\lambda_D\rightarrow\infty$) \cite{Schweigert:99,Weis:01} or the strong screening region \cite{Schweigert:99}.\\
The crystalline phase diagram of Yukawa bilayer systems is quite complicated and several arrangements have been found to be thermodynamically stable \cite{Mitchell:98,Zuzic:00,Zahn:99,Bechinger:02,Goldoni:96,Fontecha:08,Valtchinov:97,Messina:03,Totsuji:97,Schweigert:99,Weis:01}. Such complex phase diagram is interesting from an experimental point of view, in colloidal and plasma physics, and from theoretical points of view for studies of two-dimensional melting \cite{Strandburg:88}.\\
In this letter, by using Metropolis Monte Carlo simulations, we study the structure and properties of the crystalline phases of a bilayer of point particles interacting via a repulsive Yukawa potential with large $\lambda_D$ (weak screening region).\\
{\bf Methods and model} -  The system consists of $N$ point particles evenly distributed in two parallel layers separated by a distance $h$. Since this work adresses the properties of the crystalline phase, to avoid irrelevant finite size effects due to the shape of the simulation box and to permit transitions between crystalline phases, it is necessary to allow the shape of the box to change. The Monte Carlo simulations are performed at constant $N$, $T$, $h$ and $A$, where $A=L_xL_y\sin\gamma$ is the surface of the oblique simulation box where the angle $\gamma$ is variable ; $L_x$ and $L_y$ evolve such as the area $A$ remains constant. A trial move of the shape of the simulation box is done every MC-cycle. The density of particles in each layer is $\rho=N/2A$  ;  the Wigner-Seitz radius $a$ is defined by $\pi\rho a^2=1$. To fulfill electroneutrality and to achieve consistency between the Coulomb and Yukawa one component plasma each layer carries a neutralizing background with a surface charge density $\sigma=-NQ/2A$ (see ref.\cite{Mazars:07} for more details). In plasma physics, such anisotropic neutralizing backgrounds are responsible of sheath confining potentials that are often taken as parabolic potentials \cite{Tomme:00,Fortov:05}. \\
In all computations presented in this letter $h=1.0$, $Q=14$ and $\lambda_D=10\mbox{ }h$ \cite{note1} ; for a such large value of the Debye screening length $\lambda_D$, the Yukawa interaction is long ranged and the energy of the system must be computed with the Ewald method as defined in ref.\cite{Mazars:07} ; direct truncation may lead to severe bias of the MC sampling \cite{Salin:02,Mazars:07}.\\
The total energy of the bilayer of Yukawa particles in the background defined above is given by
\begin{equation}
\label{Etot}
\begin{array}{ll}
\displaystyle E&\displaystyle =\frac{1}{2}\sum_i\sum_{i\neq j}V(r_{ij})+E_B\\
&\displaystyle =E_{intra}+E_{inter}
\end{array}
\end{equation}
where $E_B$ is the sum of the energy of the particle-background interactions and the background self energy. In Eq.(\ref{Etot}), the energy is split into  intralayer $E_{intra}$ and interlayer $E_{inter}$ energies. The number of particles in the bilayer is $N=2048$ (1024 particles per layer) and, after equilibration, averages are accumulated for $5\times 10^{4}$ - $2\times 10^{5}$ trial moves per particle. A few computations have been done with $N=512$ and $1058$.\\
The intralayer $g_{11}$ and interlayer $g_{12}$ center to center correlation functions are defined by 
\begin{equation}
\label{correl1}
\begin{array}{ll}
\displaystyle g_{11}(s) &\displaystyle =\frac{1}{4\pi s\rho N_0}\Big<\sum_{i\in L_{1}}\sum_{j\in L_{1},j\neq i} \delta(s-\mid\bm{s}_{ij}\mid)\\
&\\
&\displaystyle +\sum_{i\in L_{2}}\sum_{j\in L_{2},j\neq i}\delta(s-\mid\bm{s}_{ij}\mid)\Big>\\
&\\
\displaystyle g_{12}(s) &\displaystyle = \frac{1}{2\pi s\rho N_0}\Big<\sum_{i\in L_{1}}\sum_{j\in L_{2}}\delta(s-\mid \bm{s}_{ij}\mid)\Big>
\end{array}
\end{equation}
\begin{table}
\caption{\label{table1a} Bond orientational order parameters for perfect crystal phases III, IVA and V. We denote by $\bm{b}_1$ and $\bm{b}_2$ the two primitive vectors of the two dimensional primitive cell. For all three phases, $\mid \bm{b}_1\mid = \mid \bm{b}_2\mid=b_0$ ; Phase IVA corresponds to two staggered two dimensional rhombic lattices with $\bm{b}_1.\bm{b}_2 =b_0^2 \cos\alpha$, Phase III to staggered square lattices with $\alpha=\pi/2$ and Phase V to staggered hexagonal lattices with $\alpha=\pi/3$. $N_b$ is the number of nearest neighbors in each perfect crystal phases and $\Psi_n^0$ are the values of the bond orientational parameters for each perfect crystal phases.}
\begin{center}
\begin{tabular}{cccc}
\hline
Phases & III & IVA ($\alpha$) & V \\
\hline
&&&\\
$N_b$ & 4 & 6 & 6 \\
&&&\\
$\Psi_4^0$ & 1 & $ \frac{1}{3}\mid 1+ 2\cos(2\alpha)\mid$ & 0\\
&&&\\
$\Psi_6^0$ & 0 & $ \frac{1}{3}\mid 1- 2\cos(3\alpha)\mid$  & 1\\
&&&\\
$\Psi_8^0$ & 1 & $ \frac{1}{3}\mid 1+ 2\cos(4\alpha)\mid$  & 0\\
&&&\\
$\Psi_{12}^0$ & 1 & $ \frac{1}{3}\mid 1+ 2\cos(6\alpha)\mid$ & 1\\
&&&\\
\hline
\end{tabular}
\end{center}
\end{table}
\begin{figure}
\centerline{\includegraphics[width=3.5in]{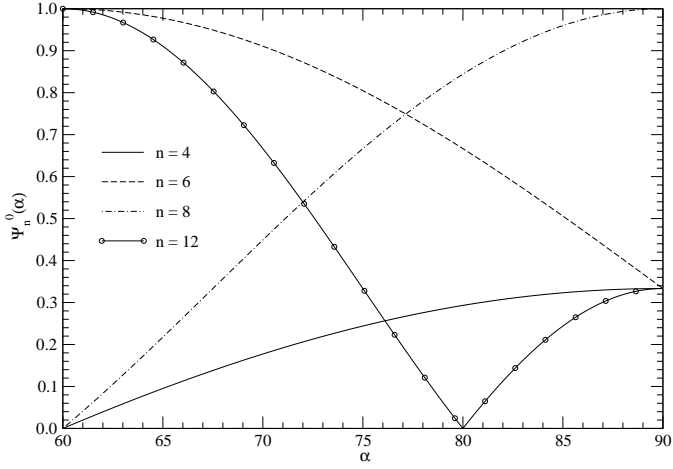}}
\caption{{\bf } Bond orientational order parameters for crystalline phase IVA as function of $\alpha$. }
\label{Fig1a}
\end{figure}
\begin{table*}
\caption{\label{table1} Simulation results for several densities for $N=2048$, $Q=14$ and $\lambda_D=10\mbox{ }h$ ($h=1.0$). Phases refer to the conventional labelling of crystalline phases developed for plasma bilayers (see for instance refs.\cite{Mitchell:98,Goldoni:96,Monarkha:03}). $\rho$ is the surface density in each layer ; $\beta U/N$ is the average energy per particle ; $\beta <E_{inter}>/N$ is the average interlayer energy per particle ;  $<\Psi_n>$ and $\chi_n$ are respectively the average bond orientational order parameters  and the susceptibilities given by Eqs.(\ref{BOOP},\ref{SUSC}) and $\alpha_g$ is the angle between the two basic primitive vectors of the two-dimensional lattices, computed from the location of peaks in the intralayer correlation functions $g_{11}(s)$ (see text and Fig.\ref{Fig4}).}
\begin{tabular}{ccccccccccc}
\hline
$\pi\rho$ &$\beta U/N$ & $\beta< E_{inter}>/N$ & $<\Psi_4>$ & $<\Psi_6>$ & $<\Psi_8>$ & $<\Psi_{12}>$ & $\chi_4$ & $\chi_6$   & $\alpha_g$ & Phases \\
\hline
0.2   & -78.52(6)   &  4.78(8)    & 0.05(5)   & 0.010(8)  & 0.05(6) & 0.011(8) & 4.76 & 0.12 & - & F/III\\
0.5   & -101.44(3) & 40.62(2)   & 0.317(2) & 0.009(6)  & 0.73(1) & 0.10(1) & 0.01 & 0.07  & $90^o$ & III \\
0.75 & -108.26(3) &  68.12(2)  & 0.318(2) & 0.008(6)  & 0.76(1) & 0.114(8) & 0.01 &  0.08 & $90^o$ & III \\
1.25 & -109.52(3) & 121.30(2) & 0.276(3) & 0.09(2)    & 0.58(2) & 0.09(7) & 0.02 & 0.95  & $88^o$ & III \\
1.35 & -108.60(3) & 131.82(2) & 0.23(1)   & 0.29(6)    & 0.40(4) & 0.03(1) & 0.22 & 8.4  & $87^o$ & IVA \\
1.4   & -108.04(3) & 137.09(2) & 0.210(7) & 0.40(3)    & 0.32(2) & 0.02(1) & 0.11  & 1.79 & $75.8^o$ &IVA \\
1.5   & -106.72(3) & 147.57(2) & 0.164(5) & 0.56(2)   & 0.16(2)  & 0.06(2) & 0.07  & 0.71 & $71.1^o$& IVA \\
1.713& -103.15(3)& 169.69(2) & 0.07(5) & 0.77(6)     & 0.02(1) & 0.4(1) & 4.44 & 7.16  & - & IVA/V\\
1.725& -102.94(4)& 171.09(2) & 0.05(4) & 0.86(6)     & 0.08(7) & 0.54(2) & 0.03  & 0.09  & $62.8^o$ & IVA/V\\
2.0   & -96.94(3) & 198.96(2) & 0.004(3) & 0.875(6)   & 0.07(5) & 0.59(2) & 0.02 & 0.06  &  $60^o$ & V \\
\hline
\end{tabular}
\end{table*}
\begin{figure}
\centerline{\includegraphics[width=3.5in]{Fig2_Mazars_G21396.eps}}
\caption{{\bf } Bond orientational order parameters as functions of the surface density for $N=2048$, $\lambda_D=10\mbox{ }h$ and $Q=14$. }
\label{Fig1}
\end{figure}
\begin{figure}
\centerline{\includegraphics[width=3.5in]{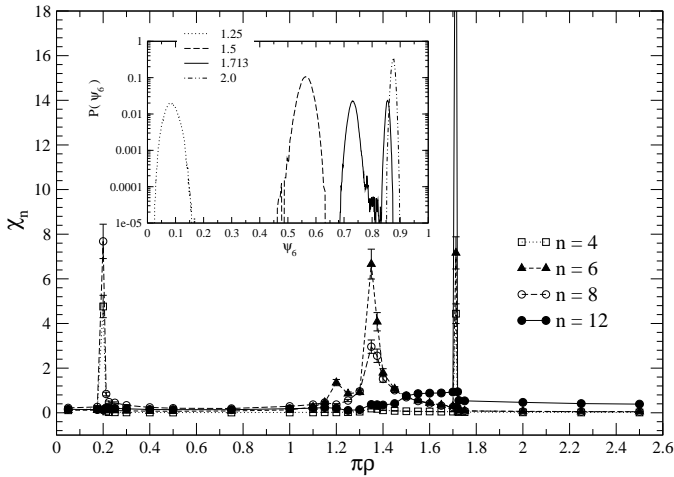}}
\caption{{\bf } Susceptibilities as functions of the surface density for $N=2048$, $\lambda_D=10\mbox{ }h$ and $Q=14$. For $\pi\rho=1.713$, we have $\chi_{12}=32.0\pm4.0$. Probability distributions for $\Psi_6$ at $\pi\rho = 1.25,$ 1.5, 1.713 and 2.0 are given in the inset.}
\label{Fig2}
\end{figure}
{\bf Bond orientational order parameters}  - For a two dimensional lattice, with $N_b$ the number of nearest neighbors defined by Voronoi constructions \cite{Collins:72}, the bond orientational order parameters are defined by
\begin{equation}
\label{Psi_0}
\displaystyle \Psi_n^0 =  \mbox{\LARGE$\mid$} \frac{1}{N_{b}}\sum_{j=1}^{N_{b}}\exp(i n\theta_{j}) \mbox{\LARGE$\mid$}
\end{equation}
where $\theta_{j}$ is the angle between the bond vector, made by a vertex of the lattice and its $j^{th}$ nearest neigbor, and an arbitrary direction. In Table \ref{table1a}, we give the values of the bond orientational order parameter of the two dimensional crystal phases (III, IVA and V) for $n=4,6,8$ and 12 ; by symmetry of the crystal phases, $\Psi_n^0$ is null for all $n$ odd. In the present work, only the three crystalline phases III, IVA and V with $\mid \bm{b}_1\mid = \mid \bm{b}_2\mid$ have been found to be stable ; the rectangular phases I and II for which $\mid \bm{b}_1\mid \neq \mid \bm{b}_2\mid$ is found in the phase diagram at $T=0$ \cite{Messina:03} are replaced by a fluid phase at the temperature considered in the present work (see below). On Fig.\ref{Fig1a}, we represent $\Psi_n^0$ as function of $\alpha$. One should note that for the perfect square crystal phase III, the Voronoi construction is degenerate : more than three Voronoi cells share a given vertex. For phase III, the Voronoi construction gives $N_b=4$ and the values of $\Psi_n^0$ have to be computed with the four neighbors.\\
As shown on Table \ref{table1a} and Fig.\ref{Fig1a}, the value of $\Psi_n^0(\pi/2 ; \mbox{III})$ for $n=4,6,12$ are not equal to $\Psi_n^0(\alpha\rightarrow \pi/2 ; \mbox{IVA})=1/3$, because the number of nearest neighbors changes from $N_b=4$, for $\alpha=\pi/2$ in phase III, to $N_b=6$ for $\alpha\rightarrow\pi/2$ in phase IVA. Therefore, the bond orientational order parameter $\Psi_4$ is not a convenient choice for the description of phases III and IVA. However, $\Psi_8$ is continuous for III$\rightarrow$IVA and one has $\Psi_8^0(\alpha\rightarrow \pi/2 ; \mbox{IVA})=\Psi_8^0(\pi/2 ; \mbox{III})=1$.\\ 
The average values of bond orientational order parameters in the MC samplings are defined by 
\begin{equation}
\label{BOOP}
\displaystyle <\Psi_n>=\frac{1}{N} \Big< \mbox{\LARGE$\mid$} \sum_i\frac{1}{N_{b,i}}\sum_{j=1}^{N_{b,i}}\exp(in\theta_{ij}) \mbox{\LARGE$\mid$} \Big>
\end{equation}
where $\theta_{ij}$ is the angle between the interparticle vector $\bm{r}_{ij}$ and an arbitrary fixed direction. The number of nearest neighbors $N_{b,i}$ of particle $i$ is determined by a Voronoi construction. The bond-orientational susceptibilities are defined as
\begin{equation}
\label{SUSC}
\displaystyle \chi_n = N(<\Psi_n^2>-<\Psi_n>^2)
\end{equation}
{\bf Results and discussion} - On Fig.\ref{Fig1}, we show the bond orientational order parameters and on Fig.\ref{Fig2}, the susceptibilities for $n=4,6,8$ and 12 ; several results for the different thermodynamical states are reported in TABLE \ref{table1}. The 'discontinuities' of $<\Psi_4>$ and $<\Psi_8>$ at $\pi\rho\simeq 0.2$ and of $<\Psi_6>$, $<\Psi_8>$ and $<\Psi_{12}>$ at $\pi\rho\simeq 1.7$ are characteristic of discontinuous first order phase transitions (the susceptibilities $\chi_4$, $\chi_6$, $\chi_8$ and $\chi_{12}$ exhibit also discontinuities at these densities). The transition at $\pi\rho\simeq 0.2$ corresponds to the transition between a disordered fluid phase and the crystalline phase III, while the transition at $\pi\rho\simeq 1.7$ is the transition between crystalline phases IVA and V (see Table \ref{table1}). \\
As outlined before, a perfect ordered square lattice is degenerate for the Voronoi construction, therefore, even very small fluctuations in position of particles will reduce greatly the number of Voronoi cells with four sides ({\it i.e.} at $T\neq 0$). Thus, the values of  $<\Psi_4>$ in phase III, computed with Eq.(\ref{BOOP}), 
never exceed $1/3$, as explained and shown on Fig.\ref{Fig1a} ; in the phase III, due to the fluctuations, each particle has on average more than four nearest neighbors. On Fig.\ref{Fig3}(a), we give a snapshot of the bilayer for $\pi\rho=0.75$ ;  for this configuration, the instantaneous bond orientational parameter are : $\Psi_4 = 0.32$, $\Psi_6 = 0.01$, $\Psi_8 = 0.74$ and $\Psi_{12} = 0.11$. Since a lattice with a square symmetry obviously has also a eight fold symmetry, the values of $<\Psi_8>$ are quite large : a moderate value of  $<\Psi_4>$ and a high value of $<\Psi_8>$ are representative of phase III (see Tables \ref{table1a}-\ref{table1} and Figs.\ref{Fig1a}-\ref{Fig1}). On Fig.\ref{Fig3}(b), we give a snapshot of the bilayer for $\pi\rho=1.35$ ;  for this configuration, the instantaneous bond orientational parameter are : $\Psi_4 = 0.23$, $\Psi_6 = 0.32$, $\Psi_8 = 0.37$ and $\Psi_{12} = 0.05$.\\
The increase of $\chi_6$ and $\chi_8$ at $\pi\rho\simeq 1.35$, with no discontinuity for $<\Psi_6>$ and $<\Psi_8>$ is associated with a continuous second order phase transition. This second order phase transition is the transition between the crystalline phases III and IVA.   \\
From the values of the order parameters reported on Fig.\ref{Fig1}, the bilayer is in a fluid phase for $\pi\rho < 0.2$, in a staggered square crystalline state (phase III) for $0.25\leq \pi\rho \leq 1.35$ ; in a rhombic crystal phase (phase IVA) for $1.35\leq \pi\rho \leq 1.7$ and in a triangular-hexagonal state (phase V) for $\pi\rho >1.7 $. The densities $\pi\rho\simeq 0.2-0.225$ and $\pi\rho\simeq 1.713$ are in the coexistence regions of the first order phase transitions. For $\pi\rho > 0.25$, this phase diagram agrees very well with the one computed in ref.\cite{Messina:03} for $T=0$ ; however, at small density, $\pi\rho<0.2$, the crystalline phases I and II, found in ref.\cite{Messina:03} are replaced by a disordered fluid phase.\\ 
It is interesting to note that $<\Psi_n>$ varies smoothly with the density over all the crystalline phase IVA ($1.35\leq \pi\rho \leq 1.7$). As shown below, in the analysis of the correlation functions, this smooth variation of the order parameters is associated with a smooth variation of the shape of the primitive cell for phase IVA ; this is also in agreement with the analysis performed in ref.\cite{Messina:03}.\\
In ref.\cite{Schweigert:99}, a melting criterion based on the values of the bond orientational order parameters $<\Psi_4>$ or $<\Psi_6>$ (noted $G_{\theta}$ in \cite{Schweigert:99}) is proposed : it is argued that melting occurs when $<\Psi_4>$ or $<\Psi_6>\simeq 0.45$. This criterion is clearly incorrect to describe the thermodynamical stability of phases III and IVA (cf. Fig.\ref{Fig1} and Fig.\ref{Fig3}). Finite size effects ($288<N<780$), poor sampling ($2\times 10^{3}$ trial moves per particle) and simulation box with fixed shape are certainly responsible for the quite large value of $G_{\theta}$ observed in ref.\cite{Schweigert:99} in comparison with the computation done for perfect crystal phases III and IVA reported in Table \ref{table1a} and Fig.\ref{Fig1a}.\\
A few MC histograms $P(\Psi_6)$, normalized to 1, are given in inset of Fig.\ref{Fig2} for $\pi\rho=1.25$, 1.5, 1.713 and 2.0. For $\pi\rho=1.713$, the histogram is double peaked since the system is in the coexistence region for this density. Single peaked histograms are quite well fitted by $P(\Psi_n)=P_{0,n}(\rho)\exp(-N(\Psi_n-<\Psi_n>)^2/2\chi_n)$.\\
\begin{figure}
\centerline{(a)\includegraphics[width=3.3in]{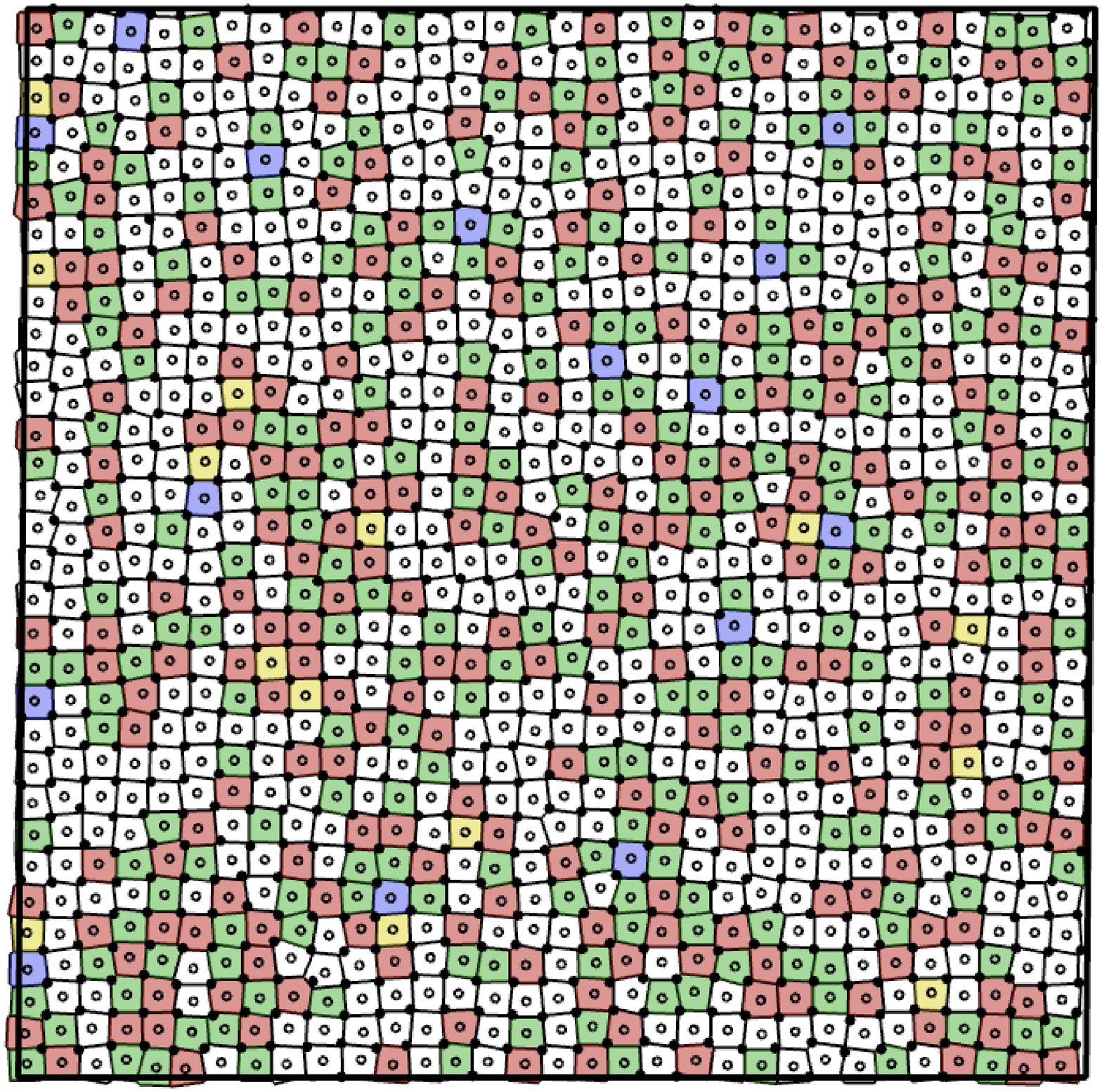}} \centerline{(b)\includegraphics[width=3.3in]{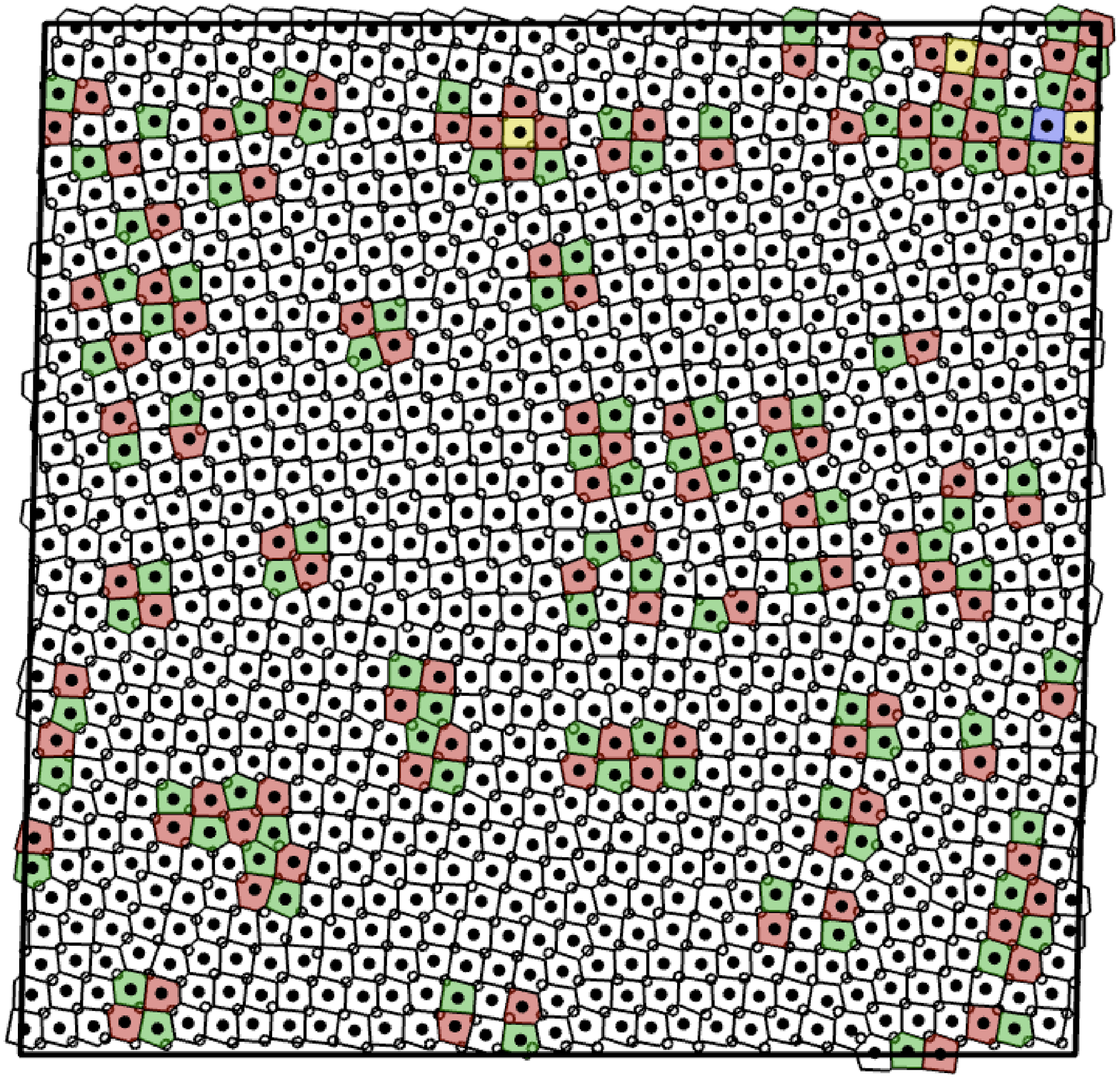}}
\caption{{\bf }(color online) Snapshots of bilayer systems with Voronoi construction for one layer, $N=2048$. Particles belonging to different layer are represented by open and solid circles. Voronoi cells with four sides are represented in yellow, those with five sides are represented  in green, those with six sides in white, those with seven sides in red and those with eight sides in blue. The sides of the simulation box are represented by thick black lines, periodic boundary conditions are applied ; (a) $\pi\rho=0.75$ ($\gamma= 89.5^o$) and (b) $\pi\rho=1.35$ ($\gamma=88.9^o$).}
\label{Fig3}
\end{figure}
\begin{figure}
\centerline{\includegraphics[width=3.5in]{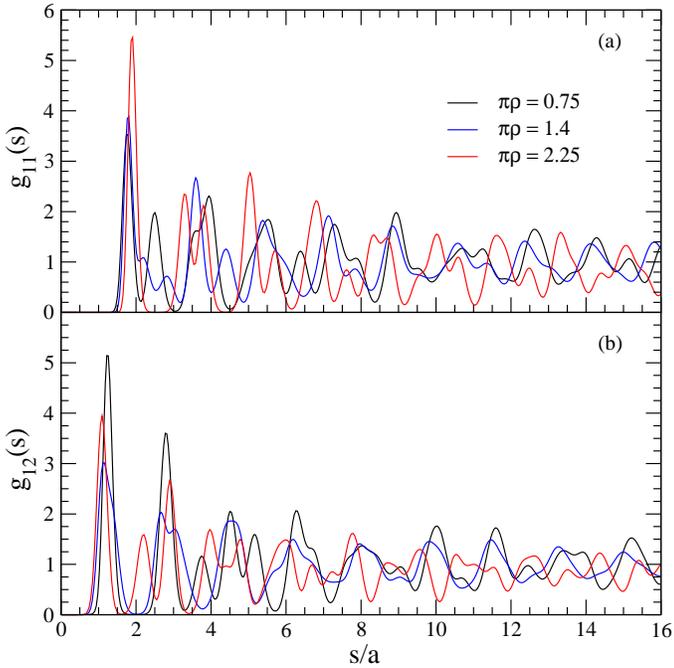}}
\caption{{\bf }(color online) Intralayer $g_{11}(s)$ and interlayer $g_{12}(s)$ center-to-center correlation functions for the three crystalline phases. The correlation functions in black correspond to crystalline phase III (staggered square lattices) ; in blue to phase IVA (staggered rhombic lattices) and in red to phase V (staggered hexagonal lattices).}
\label{Fig4}
\end{figure}
On Fig.\ref{Fig4}, we show the intralayer $g_{11}(s)$ and interlayer $g_{12}(s)$ center-to-center correlation functions for the three crystalline phases III, IVA and V. From the values of $g_{11}(s)$ we can determine the parameters and the structure of the primitive cell of the two dimensional Bravais lattice in each layer. In crystalline phases, the particles fluctuate around their equilibrium position defined by the sites of the two dimensional lattice, thus $g_{11}(s)$ may be accurately fitted by a sum of gaussian functions of the form $g_n(s)=g_{0n}\exp(-(s-S_n)^2/2\sigma_n^2)$ where $S_n$ is the location of the n$^{th}$ peak. For the rhombic primitive cell (phase IVA), if we denote by $\bm{b}_1$ and $\bm{b}_2$ the two primitive vectors of the two dimensional primitive cell and by $\alpha$ the angle between these two primitive vectors, then the first three peaks are located at $S_1=b_0$, $S_2=b_0\sqrt{2(1-\cos\alpha_g)}$ and $S_3=b_0\sqrt{2(1+\cos\alpha_g)}$ with $b_0=\mid\bm{b}_1\mid=\mid\bm{b}_2\mid$. For densities $1.25\leq \pi\rho \leq 1.6$, a fit of the first three peaks of $g_{11}(s)$ by a sum of three independent gaussian functions gives $b_0=1.80\mbox{ }a$ and $90^o<\alpha_g<60^o$ (some values of $\alpha_g$ are reported on Table \ref{table1}). For $1.35\leq \pi\rho\leq 1.45$, $<\Psi_{12}>$ is almost null while $<\Psi_6>$ and $<\Psi_8>$ are almost equal, according to the computations reported on Fig.\ref{Fig1a} the angle of the rhombic primitive cell is $\alpha\simeq 80^o$ and agree well with the value of $\alpha_g$ computed from correlation functions (cf. Table \ref{table1}).\\
For phases III and V, a similar analysis of  $g_{11}(s)$ can be done. For phase III, we found $b_0=1.80\mbox{ }a$ and $\alpha_g=90^o$ (the peaks located at $S_2$ and $S_3$ in phase IVA merge into a single peak - the second one - in phase III, see Fig.\ref{Fig4}(a)) ; for phase V, we have $b_0=1.90\mbox{ } a$ and $\alpha_g=60^o$ (the peaks located at $S_1$ and $S_2$ in phase IVA merge into a single peak).\\
The analysis of the peaks of $g_{12}(s)$ can be done in a similar way. For the phase III (staggered square lattices),  the first two peaks of $g_{12}(s)$ must be located at $S_1'=b_0/\sqrt{2}$ and $S_2'=b_0\sqrt{5/2}$. For $\pi\rho<1.3$, these values are well reproduced. For instance, for $\pi\rho=1.0$, fitting the peaks by sums of gaussian functions, from $g_{11}(s)$ we found $b_0=1.77\mbox{ }a$ and from $g_{12}(s)$ we have $S_1'=1.24\mbox{ } a$ and $S_2'=2.80\mbox{ } a$. For $\pi\rho > 1.35$, the system is in phase IVA and  the first two peaks of $g_{12}(s)$, observed in phase III, separate each one into two peaks (see the blue curve in Fig.\ref{Fig4}(b)).\\
On Fig.\ref{Fig5}, we show snapshots of instantaneous configurations of the bilayer system with the Voronoi construction for one layer. On these figures, Voronoi cells with different number of sides are represented with different colors. The irregular hexagonal Voronoi cells are typical of the rhombic primitive cell of phase IVA. As may be seen on these figures, particles belonging to one layer are mainly located on the edge of the Voronoi cells of the particles belonging to the other layer.\\
\begin{figure}
(a)\includegraphics[width=3.3in]{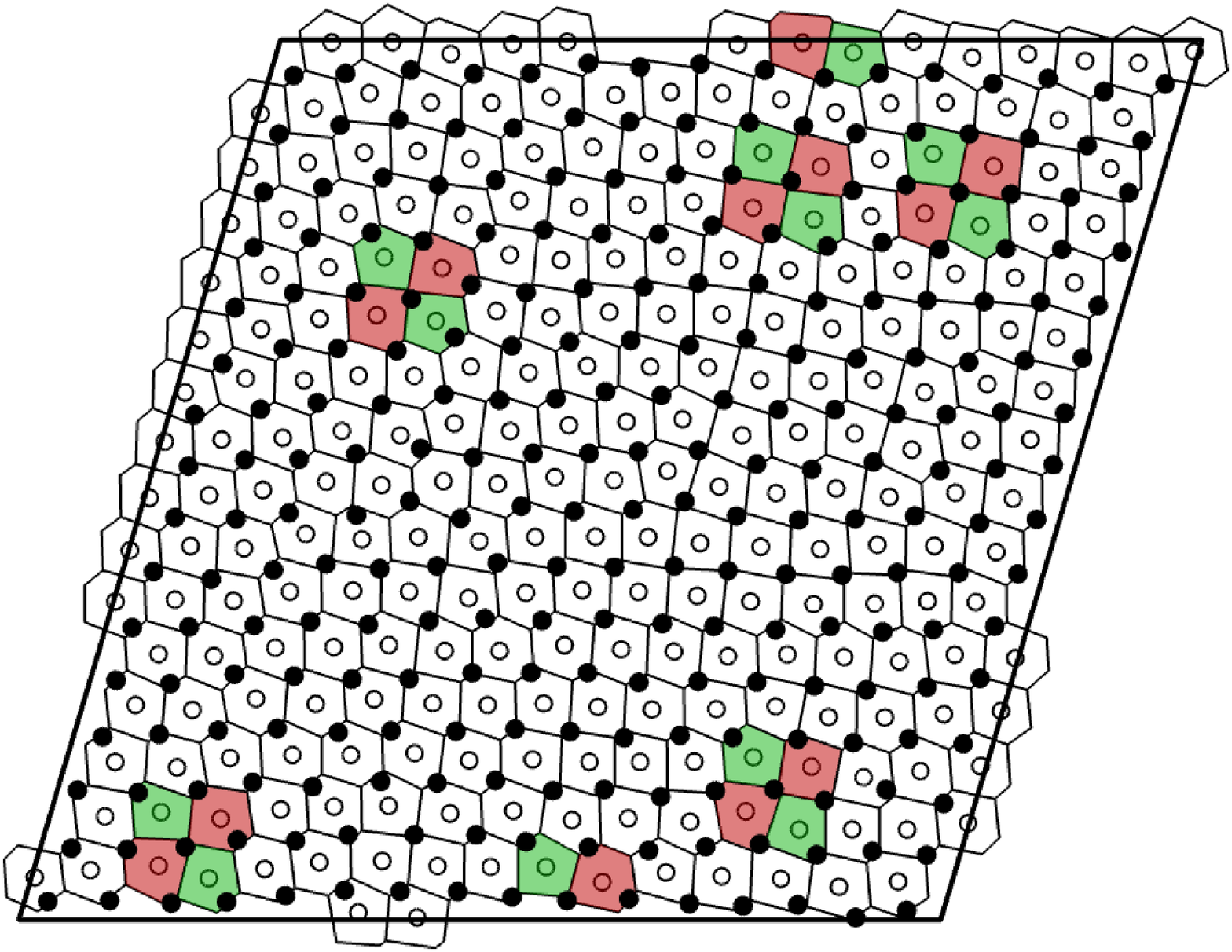}\\(b)\includegraphics[width=3.2in]{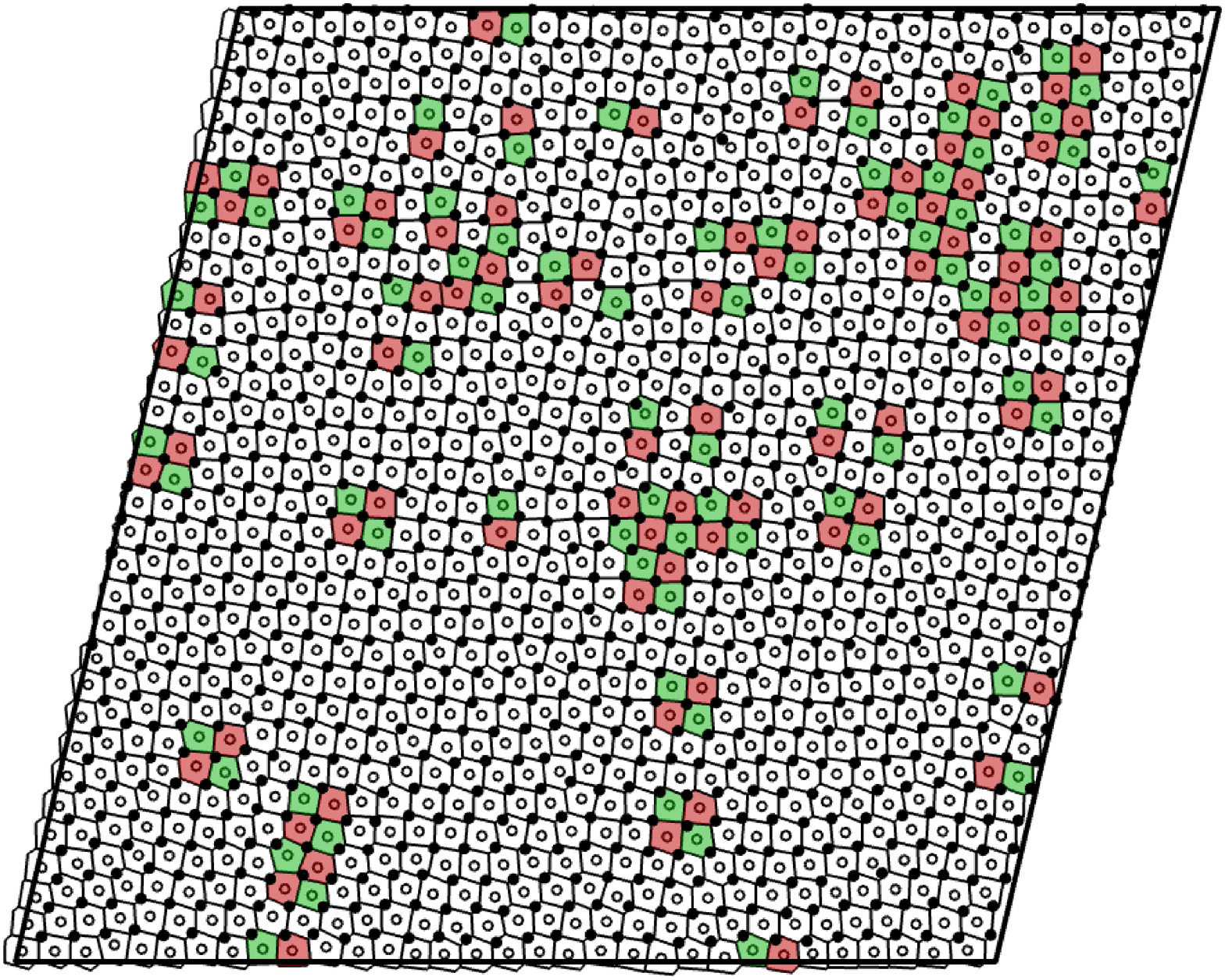}
\caption{{\bf } (color online)  Same as Fig.\ref{Fig3} with $\pi\rho=1.4$ (a)  $N=512$ and (b) $N=2048$. The rhombic shape of the simulation box results from the MC trial moves of the shape of the box : (a) $\gamma=73.8^o$ ($<\gamma>=74.4^o\pm1.1^o$) and (b) $\gamma=77.4^o$ ($<\gamma>=76.6^o\pm0.6^o$). Whereas $<\gamma>$ is strongly correlated to the crystalline order of the phase, it may and can not be used as an order parameter because of defects (see for instance Fig.\ref{Fig3} and TABLE \ref{table1}).}
\label{Fig5}
\end{figure}
As it appears on Fig.\ref{Fig5}, the main influence of the finite size effects is to reduce the number of dislocations per surface area. This effect improves the stability of the crystalline phases \cite{Strandburg:88}. It is also worthwhile to note that, for the temperature considered in this letter, dislocations with five-fold and seven-fold symmetries are grouped together in 'neutral' clusters (pairs, quartets, etc.) ; larger clusters are observed in larger systems. Similar results have been observed experimentally on monolayers of strongly coupled dusty plasma monolayers \cite{Feng:08}.\\
All computations presented here are done in the weak screening region of Yukawa potentials where the interaction is long ranged and the use of Ewald sums necessary \cite{Mazars:07,Salin:02}. The phase diagram shows three different phase transitions : two first order transitions, at $\pi\rho\simeq  0.2$ (fluid$\rightarrow$ III) and at $\pi\rho\simeq 1.7$ (IVA$\rightarrow$ V) and a second order transition at $\pi\rho\simeq 1.35$ (III$\rightarrow$ IVA). This phase diagram, except for $\pi\rho\leq 0.2$, is in very good agreement both qualitatively and quantitatively (in the numerical values of the densities of the phase coexistence \cite{note1}) with the phase diagram computed in ref.\cite{Messina:03} at $T=0$.\\

\acknowledgments
It is a pleasure to thank J.-J. Weis for a lot of interesting discussions and V. Huet for her help in the preparation of the manuscript.\\
The author acknowledges computation facilities provided by the {\it Institut du D\'eveloppement et des Ressources en Informatique Scientifique\/} (IDRIS) under project 0682104.


\begin{thebibliography}{0}
\bibitem{note1} These values of $h$ and $\lambda_D$ correspond to $\lambda=0.1$ in notations of ref.\cite{Messina:03} and to $\lambda\nu=0.1$ in notations of ref.\cite{Schweigert:99}. More precisely, the parameters $(\nu,\Gamma,\lambda)$ defined in ref.\cite{Schweigert:99} are defined in notations of the present letter as : $\nu=h\sqrt{\pi\rho}$, $\Gamma=Q^2\sqrt{\pi\rho}/k_BT$ and $\lambda=1/\lambda_D\sqrt{\pi\rho}$ ; also, in notations of ref.\cite{Messina:03}, the reduced density and the relative screening length are given respectively here by $\eta = \rho h^2$ and $\lambda=h/\lambda_D$.
\end{thebibliography}
\end{document}